# AN AUTOMATED ALGORITHM FOR APPROXIMATION OF TEMPORAL VIDEO DATA USING LINEAR BEZIER FITTING

Murtaza Ali Khan

Faculty of Information Technology
Royal University for Women, Bahrain
mkhan@ruw.edu.bh

## ABSTRACT

*This paper presents an efficient method for approximation of temporal video data using linear Bezier fitting. For a given sequence of frames, the proposed method estimates the intensity variations of each pixel in temporal dimension using linear Bezier fitting in Euclidean space. Fitting of each segment ensures upper bound of specified mean squared error. Break and fit criteria is employed to minimize the number of segments required to fit the data. The proposed method is well suitable for lossy compression of temporal video data and automates the fitting process of each pixel. Experimental results show that the proposed method yields good results both in terms of objective and subjective quality measurement parameters without causing any blocking artifacts.*

## 1. INTRODUCTION

Digital video data consists of sequence of frames (images). Each frame consists of rectangle 2-D array of pixels. 3-D *RGB* or 1-D *intensity* values in a sequence of frames are associated with each pixel. Tuple of a pixel in a frame can be considered a point in Euclidean space $R^3$ or $R^1$. Therefore, if a video consists of a sequence of *n* frames then for each pixel we have a set of values $\{p_1, p_2, pn\}$, i.e., $p_j = (R_j, G_j, B_j)$ for *RGB* or $p_j == I_j$ for *intensity*, where $1 \leq j \leq n$. Figure 1 shows *RGB* variation of a pixel in 80 frames of a video.

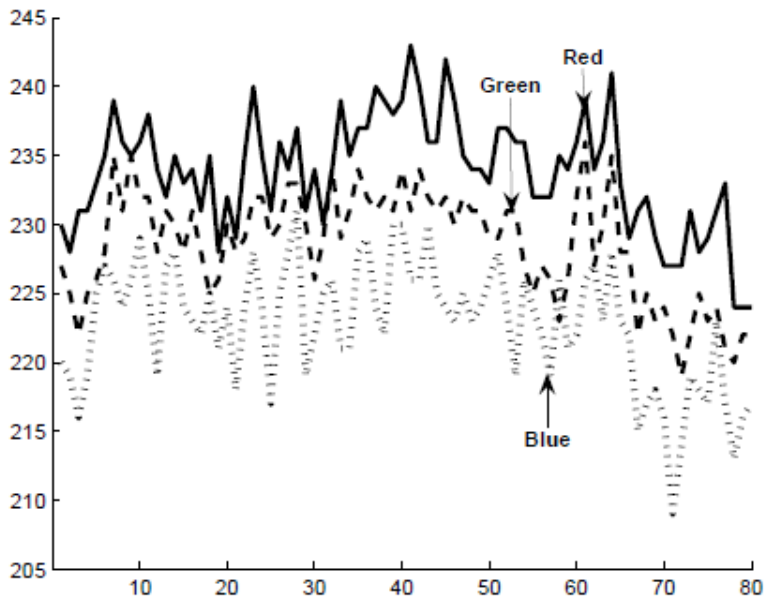

**Figure 1: RGB temporal variation of a pixel in 80 frames of a video.**

Video data contains temporal and spatial correlation. Spatial correlation of video data is reduced by predictive or transform coding techniques while temporal correlation is reduced by techniques like motion estimation. In our proposed method of video data approximation focus is on reducing temporal correlation of video data by approximating it using linear Bezier fitting (LBF). Due to large size of video data it is also desirable that approximation process is automated. In our proposed scheme only initially does the user have to set few parameters, and then rest of the fitting process is fully automated.

## 2. RELATED WORK

Approximation of data using parametric curves, particularly cubic splines is explored by many authors [1, 2, 3, 4] et al. But almost in all of these techniques fitting is for non-video data and most of these techniques require user interaction. The method presented by [5], approximates the video data using parametric line and Natural cubic spline by combining the group of pixel together as a block and the applies fitting to block. The approach we adopted in this paper is based on linear Bezier curve fitting at pixel level. Pixel level fitting yields more precise control of accuracy compare to block level fitting. A video object encoding algorithm based on the curve fitting trajectory of video object moving edges pixels is presented by [6]. The algorithm of [6] is suitable, if the objects in the video are not moving quickly such as in video conferencing or surveillance. Our method does not encode the edges of objects only; but it applies fitting to luminance/color variations of every pixel. Non-spline based temporal correlation reductions methods are based on motion estimation via translating block matching algorithms (BMAs) [7, 8, 9, 10]. In a typical BMA, a frame is divided into rectangle blocks of pixels. Then the current block (predicated block) is matched against blocks in the previous frame, for a maximum motion displacement of $w$ pixels, i.e., $\pm w$ pixels in both horizontal and vertical directions. The best match on the basis of a *mean absolute error MAE* criterion yields displacement relative to current block called motion vector. Predicated frame is approximated by blocks in reference frame and corresponding motion vectors [11, 12, 13, 14]. Rather than finding matching pixel or matching block, the proposed method adopts the different approach of fitting i.e., it approximates the change in intensity of each pixel at the fixed spatial location (without translation of block/pixel) in the set of consecutive frames by linear Bezier fitting. BMA works at block level and may cause blocking artifacts [13, 15]. Due to pixel level fitting by proposed method, precise control of accuracy and immunity from blocking artifacts is achieved. An important feature of proposed method is that it is simple and fully automated, thus well suited for video data approximation/compression.

Organization of the rest of the paper is as follows: Parametric linear Bezier model is briefly described in Sect. 3. Section 4 is the most important section and it describes the details of fitting strategy and proposed algorithm. Experiments and results are presented in Sect. 5. Section 6 gives insight view of the proposed method. Final concluding remarks are in Sect. 7.

## 3. PARAMETRIC LINEAR BEZIER

Parametric linear Bezier is essentially a straight line obtained by linear interpolation between two control points. To generate linear Bezier that interpolates $k+1$ points (intensity or color variation in $k+1$ frames), $k$ line segments are used. Equation of $j^{th}$ segment between points $p_j$ and $p_{j+1}$ can be written as follows:

$$q_j(t) = (1-t)p_j + tp_{j+1}, \quad t \in [1,0], \quad 1 \leq j \leq k, \qquad (1)$$

where $q_j(t)$ is an interpolated point between $p_j$ and $p_{j+1}$ at parameter value $t$. To generate $n$ points between $p_j$ and $p_{j+1}$ inclusive, the parameter $t$ is divided into $n$-1 intervals between 0 and 1 inclusive such that $q_j(0) = p_j$ and $q_j(1) = p_{j+1}$

## 4. FITTING STRATEGY AND ALGORITHM

In this section we will describe the strategy of proposed linear Bezier fitting algorithm to video data. Fitting process is applied to temporal data of each pixel individually. Color or luminance value of a pixel at frame $i$ is $p_i$, where $0 \leq p_i \leq 255$ and $1 \leq i \leq n$. We have to approximate the $n$ values of each pixel i.e., original data, O={$p_1, p_2,...,p_n$}, by linear Bezier fitting. As an input to algorithm the user specifies two parameters: (1) *upper limit of error* $\lambda^{lmt}$, i.e., maximum allowed mean squared error between original and fitted data, e.g., $\lambda^{lmt} = 100$ (2) *initial keypixel interval* $\Delta$, i.e., pixel after every $\Delta$ frames is taken as a keypixel, e.g., $\Delta = 12$ then set of initial keypixels is KP={$p_1, p_{13}, p_{25}, p_{37},...,p_n$} (pixel of last frame is always taken as keypixel). The fitting process divides the data into segments based on keypixels. A segment is set of all points (pixel values) between two consecutive keypixels, e.g., $S_1$={$p_1, p_2,...,p_{13}$}, $S_2$={$p_{13}, p_{14},...,p_{25}$}. Bezier interpolation is performed for each segment and $n$ interpolated values, i.e., approximated data, Q={$q_1, q_2,...,q_n$}, is obtained (common interpolated values are removed at the joint of two segments). Then we compute the mean squared error of each segment. If mean squared error of any segment is greater than $\lambda^{lmt}$ then this segment is split (replaced) by two new segments. Since a segment consists of many points, we have to select a point where this segment is going to be split. The point, where the squared distance between each point of original data and its corresponding point on approximated data is maximum, is selected as a split-point of this segment. A new keypixel $kp_{new}$ from original data is added in the set of keypixels where the error is maximum. For example if segment $S_1$ is split at $p_6$ then a new keypixel $kp_{new} = p_6$ is inserted between keypixels $p_1$ and $p_{13}$, i.e., is KP={$p_1, p_6, p_{13}, p_{25}, p_{37},...,p_n$} and two new segments {$p_1, p_2,...,p_6$} and {$p_6, p_7,...,p_{13}$} replace $S_1$. The fitting process is repeated with new set of keypixels until mean square error of each segment is equal or less than $\lambda^{lmt}$. The Algorithm 1 describes the proposed algorithm formally. Figure 2 explains how a segment splits. There is maximum error in frame 12 (top) between original and fitted (approximated) data, therefore new keypixel is added at frame 12 (bottom). Figure 3 shows fitting of intensity data of a pixel in 80 frames of a video by linear Bezier. It is evident that in addition to initial keypixels, other keypixels are inserted at arbitrary frames, where needed (due to split of segments), depends on the intensity variations of each individual pixel.

---

**Algorithm 1** Break-and-fit linear Bézier fitting, segment level error bound

---

**Require:** Points of original data: $O = \{p_1, p_2, \ldots, p_n\}$,
  Initial set of keypixels $KP = \{kp_1, kp_2, \ldots, kp_l\}$,
  Maximum allowed $MSE$ i.e., $\lambda^{lmt}$.

1: Divide $O$ into segments $S = \{S_1, S_2, \ldots, S_{l-1}\}$
2: Fit the linear Bézier to each segment, i.e., Find $Q = \{q_1, q_2, \ldots, q_n\}$
3: Find $MSE$ of each segment i.e., $\lambda_1, \lambda_2, \ldots, \lambda_{l-1}$
4: $\lambda^{max} = Max(\lambda_1, \lambda_2, \ldots, \lambda_{l-1})$, $\lambda^{max} \in j^{th}$ segment
5: **while** $\lambda^{max} > \lambda^{lmt}$ **do**
6:   Find $d^2_{max}$ of $j^{th}$ segment, $d^2_{max} = \|p_i - q_i\|^2$, $kp_{new} = p_i$
7:   Add $kp_{new}$ in $KP$
8:   Split $j^{th}$ segment into $j_1^{th}$ and $j_2^{th}$ segments
9:   Using updated $KP$ fit $j_1$, $j_2$ segments
10:  Find $\lambda$ of $j_1$ and $j_2$ segments only
11:  Find new $\lambda^{max}$, $\lambda^{max} \in j^{th}$ segment
12: **end while**

---

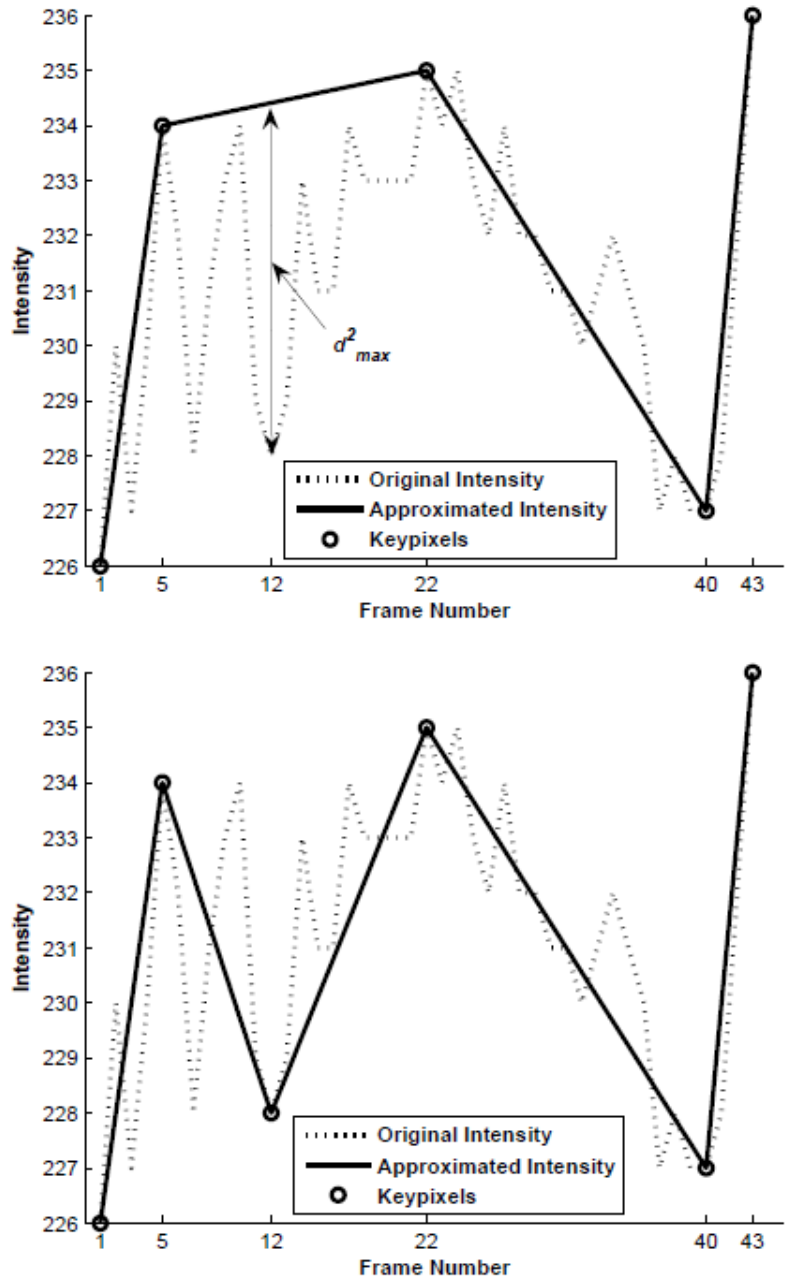

**Figure 2: Top: Max error occurs at frame 12. Bottom: a new keypixel is inserted at frame 12.**

The International Journal of Multimedia & Its Applications (IJMA), Vol.2, No.2, May 2010

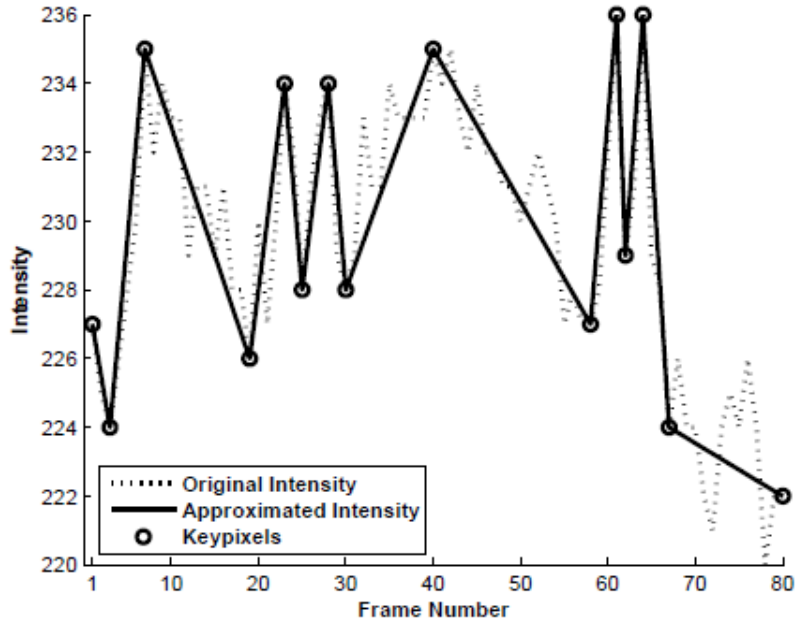

**Figure 3: Linear Bezier fitting to intensity variations of a pixel in 80 frames of a video. 80 pixel values are approximated with 15 keypixels values. Maximum Allowed *MSE*=5.**

The output data need to be saved and required to decode the approximated video is *intensity value of keypixels* and *count of interpolating points between keypixels* (*cip*). The decoding is simple linear interpolation using Eq. 1. Rather than saving *cip* separately, a better approach is to substitute intensity value of each non-keypixel by some non-occurring value/symbol in video data (e.g., -1 or 256). By counting occurrences of that value, *cip* can be determined easily. For example if there are keypixels in frame 1,6 and 13 then we can save { $p_1$, -1, -1, -1, -1, $p_6$, -1, -1, -1, -1, -1, -1, $p_{13}$}. Since -1 occurs 4 times between $p_1$ and $p_6$, it means *cip* is 4 between $p_1$ and $p_6$, exclusive. Similarly -1 occurs 6 times between $p_6$ and $p_{13}$, it means *cip* is 6 between $p_6$ and $p_{13}$, exclusive. At the decoding time approximating intensity values of non-keypixels are obtained by simple linear interpolation of intensity values of keypixels. Then -1 is substituted with these approximating intensity values of non-keypixels and approximated video is constructed. This approach reduces the entropy (data can be coded at low bitrate), intact the rectangular shape of image and does not require to save side information (e.g. in addition to reference frames, motion vectors need to be saved in BMAs).

## 5. EXPERIMENTS AND RESULTS

Extensive simulation is done on various standard test video sequences using proposed algorithm. The performance of approximated video data is evaluated in terms of *Entropy* of output data, measured in bits per pixel (bpp) and *PSNR*, measured in decibel (dB). Less *Entropy* and high *PSNR* are desirable.

$$Entropy = -\sum_{j=1}^{J} P(a_j) \log P(a_j), \qquad (2)$$

$$MSE = \frac{1}{MN}\sum_{i=1}^{M}\sum_{j=1}^{N}\left\|p_{i,j}-q_{i,j}\right\|^2, \quad (3)$$

$$PSNR = 10\log 10\left(\frac{255^2}{MSE}\right), \quad (4)$$

where $J$ is the unique number of symbols in the source, $P(a_j)$ is the probability of the occurrence of symbol $a_j$. $M$ and $N$ are width and height respectively of a frame, $p_{i,j}$ and $q_{i,j}$ are original and approximated intensities respectively of a pixel at spatial location $(i, j)$.

Table 1 gives the details of input video sequences used in simulation. *Foreman* sequence has medium temporal activity and *Football* sequence has high temporal activity. Figure 4 and Figure 5 show variation of *Entropy* and *PSNR* at various values of $\lambda^{lmt}$ for *Foreman* and *Football* video sequences respectively. Figure 6 and Figure 7 show 16$^{th}$ frame of original, DS-BMA approximation and LBF approximation for *Foreman* and *Football* video sequences respectively. Figure 8 and Figure 9 show 16$^{th}$ frame of approximated *Foreman* and *Football* video sequence respectively where non-keypixels (approximated pixels) are shown in white color.

**Table 1: Details of input video sequences used in simulation**

| Video Name | Format | Frame Size | Number of Frames |
|---|---|---|---|
| *Foreman* (*Y* Component) | SIF | 352 x 288 | 44 |
| Football (*Y* Component) | SIF | 352 x 240 | 44 |

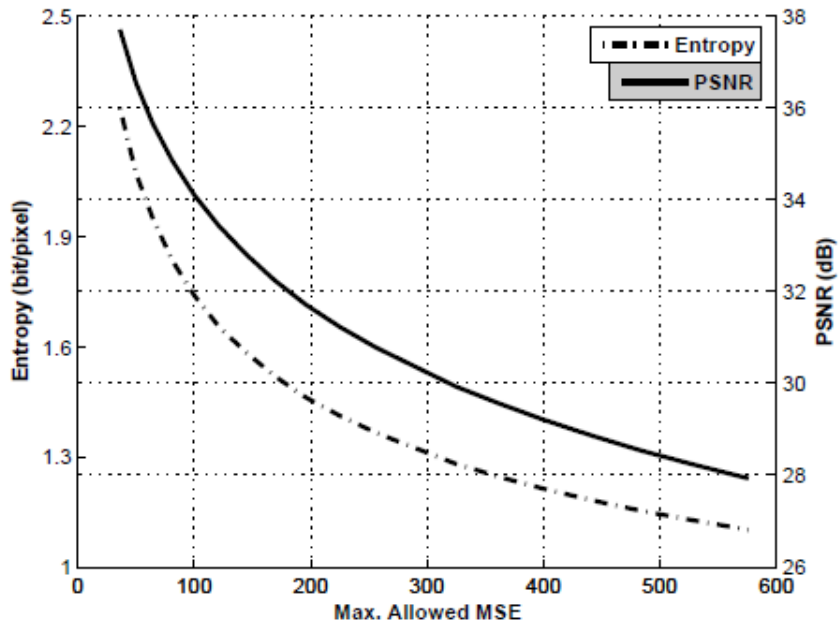

**Figure 4:** *Entropy* and *PSNR* performance of *Foreman* sequence at varying value of maximum allowed mean squared error ($\lambda^{lmt}$).

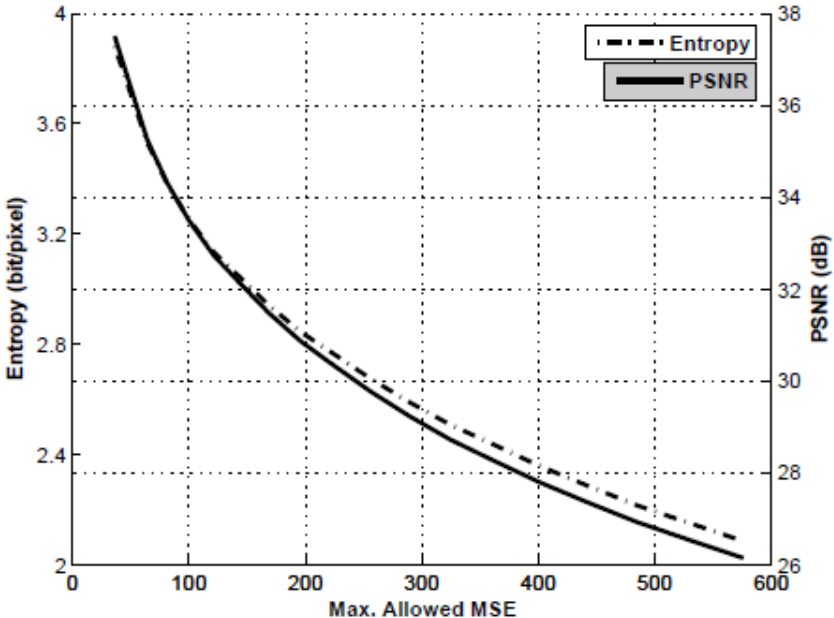

**Figure 5: Entropy and PSNR performance of *Football* sequence at varying value of maximum allowed mean squared error ($\lambda^{lmt}$).**

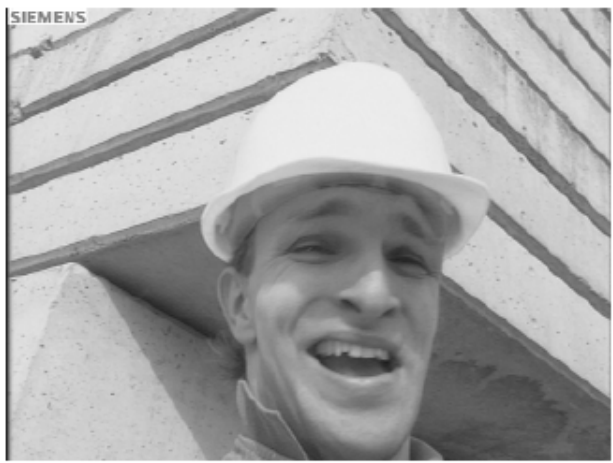

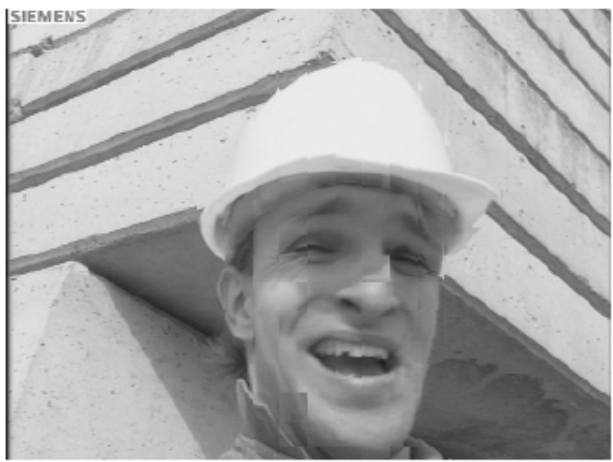

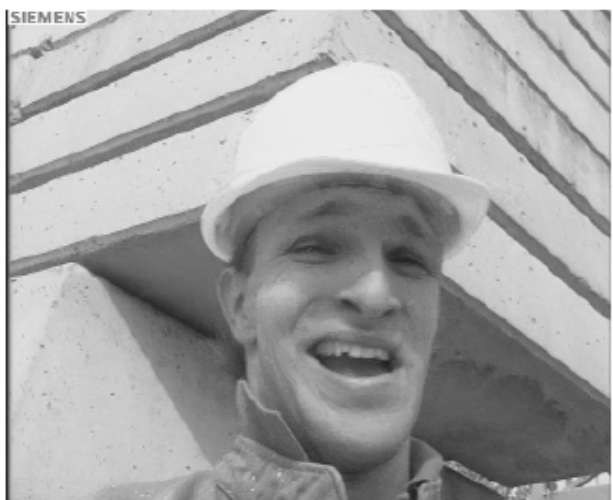

**Figure 6: 16$^{th}$ frame of *Foreman* video sequence. Original (top), DS-BMA approximation (middle), LBF approximation (bottom).**

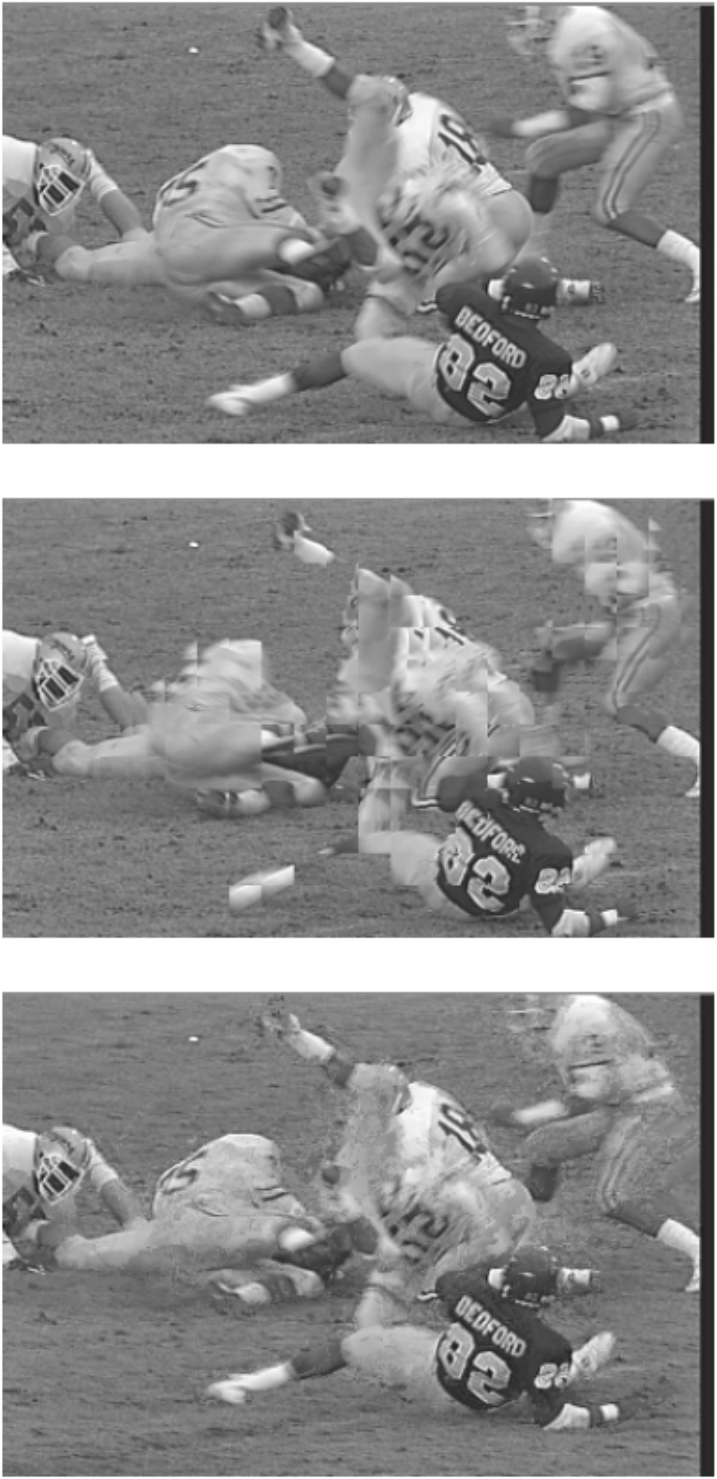

**Figure 7: 16<sup>th</sup> frame of *Football* video sequence. Original (top), DS-BMA approximation (middle), LBF approximation (bottom).**

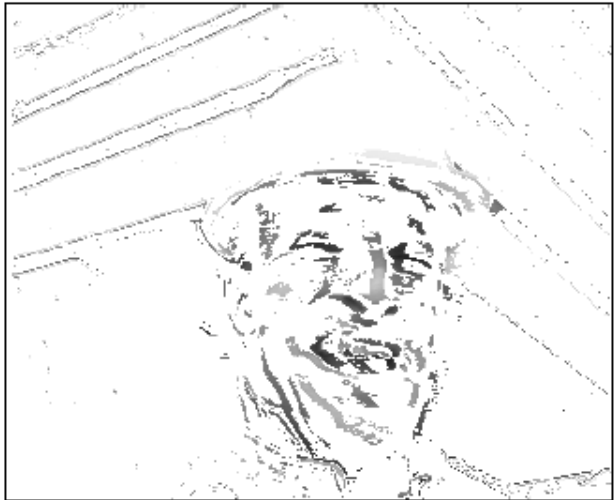

**Figure 8:** 16<sup>th</sup> frame of approximated *Foreman* video sequence by LBF. Non-keypixels (approximated pixels) are shown in white.

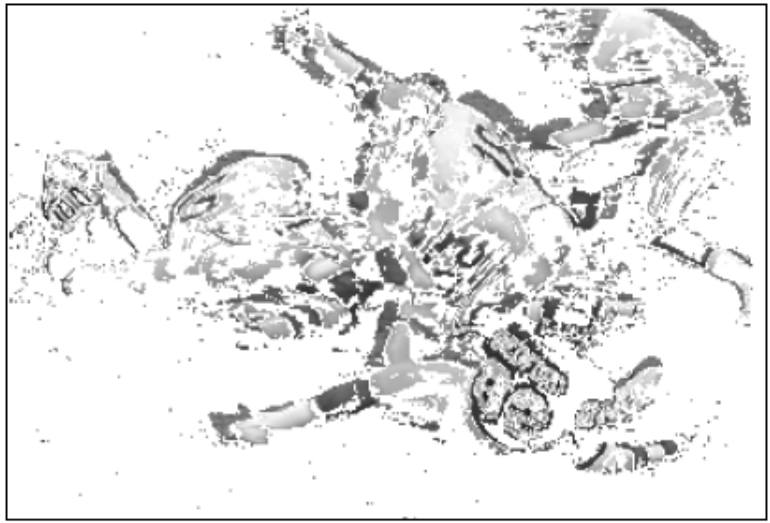

**Figure 9:** 16<sup>th</sup> frame of approximated *Football* video sequence by LBF. Non-keypixels (approximated pixels) are shown in white.

## 6. DISCUSSION

*Analysing Results*: Human visual system (HVS) is more sensitive to edges [13, 16] than arbitrary noise. Error due to block matching of BMA produces edges while error due to pixel level LBF produces arbitrary or randomly distributed error. Therefore, approximated images of BMA tend to have low subjective quality than approximated images of LBF, even though both have error. *Foreman* sequence has motion activity in the face while the background is fixed. Therefore, error in approximated images/videos is lie on face. *Football* sequence has lot of motion activity, i.e., change in intensity of pixels is rapid, both in the background and foreground. Matching blocks are scarce in reference frame and error between blocks of predicated and reference frame is higher, consequently BMA cause quite noticeable blocking artifacts. Due to pixel level fitting and break and fit strategy, LBF approximates the change in intensity of pixels relatively better, even though it also contains error.

From Figure 8 and Figure 9, where non-keypixels are shown in white color, it is evident that the LBF successfully approximates majority of the pixels. Almost all the background and considerable portions of foreground in both videos are approximated by fitting (i.e., no data need to be saved for non-keypixels).

***Reasons to choose linear Bezier***: (1) Due to large size of video data simple and efficient fitting is inevitable. Linear Bezier is most efficient and simple than other types curves e.g., quadratic/cubic Bezier curves, Natural cubic spline, B-spline, Cardinal spline etc. There is no continuity constraint in the adjacent segments of Bezier curves i.e., each linear Bezier segment can be constructed independently of other segments and abrupt changes in pixel intensity can be approximated by multiple linear Bezier segments efficiently. (2) There is no *middle control point* in linear Bezier unlike quadratic/cubic Bezier curves, further *end-control points* are in the same data range, as original data, i.e., [0-255]. Therefore, linear Bezier needs less data (low entropy) to save and is suitable for video data compression. (3) Fitting of data by linear Bezier is local whereas fitting of data using other types of splines e.g., Natural cubic spline or Cardinal spline is non-local. Local fitting means breaking of a segment into two segments in Bezier fitting requires only re-computation of two newly created segments without affecting the remaining segments. Therefore whenever a fitted segment splits due to large change in pixel intensity, the computation cost remains within acceptable limit. The locality of Bezier fitting is at the expense of less continuity. Linear Bezier is $C^0$ continuous, this result in lack of *smoothness*. But this lesser continuity does not impact the quality of reconstructed video because it depends on *closeness* of fit to original data rather than *smoothness* of fit. Smoothness of fit is desirable feature in applications like font design where smooth curves are more pleasing.

***Selection of keypixels:*** In the described method, optimal fitting is achieved for initial given set of keypixels at certain regular intervals. We have not used any model to give best set of initial keypixels and consequently division of segments might not be optimal from fitting perspective. In conventional data fitting methods corner detection is frequently used to given initial set of break points. But for video data detecting corners for each pixel data is computationally not feasible. This will also result in different initial set of break points (keypixels) for each pixel. Giving same set of keypixels at regular intervals for each pixel gives another nice feature that is encoded video can be decoded between two known intervals.

***Fitting in higher dimension:*** The proposed method can approximate video data of any dimension e.g. 1-D *luminance*, *chrominance* signals or 3-D *RGB*, $YC_bC_r$, *HSV* signals. Theoretically the proposed method can work for *N*-D points i.e. in Euclidean space $R^N$.

***Approximation for compression:*** The main purpose of approximation of temporal video data is lossy compression. From Figure 4 and Figure 5 it can be seen that the entropy of output data is very low, it means it can be coded with less bitrate. Further reduction in entropy or bitrate can be achieved by spatial compression, quantization, arithmetic coding, run-length coding etc.

***Intensity variations without motion:*** Motion essentially changes intensity of pixels in temporal dimension. But there are situations where intensity of pixels change without motion e.g., increase or decrease in the intensity of light (artificially or naturally) in the scene without motion of any object. Since fitting approximates the change in intensity either it is due to motion or without motion, therefore it works in both cases. But if a model is based on motion then if the intensity change without motion then model can not find matching object/block of same intensity in the previous frame, even though the same object/blocks exists at the same position but with different intensity. This also suggests the usefulness of proposed strategy.

# 7. CONCLUSIONS

We presented a simple method for approximation of temporal video data by linear Bezier fitting. The proposed method is suitable for lossy compression of temporal video data. The method fully automates the fitting process and each pixel data is processed individually and independently. The method can be applied to temporal video data of any dimension (e.g., *RGB* and intensity etc.). Experimental results show that the proposed method yields good results both in terms of objective and subjective quality measurement parameters i.e. *Entropy/PSNR* and *human visual acceptance* without causing any blocking artifacts.

**Authors**

Dr. Murtaza Khan received his Ph.D., degree from Keio University, Japan in 2008. His Master of Science degree in Computer Science was obtained from K.F.U.P.M, KSA in 2001. His areas of research include: image and video processing, data compression, spline fitting, and computer animation. He has published several papers in referred journals and international conferences. Dr. Khan received many awards including Excellence in Research Award for year 2001-02, ICS department, K.F.U.P.M, Yoshida Scholarship Award for doctorate studies 2004-07, and research grant from Keio Leading-edge Laboratory of Science and Technology for three consecutive years from 2005-07. Dr. Khan is currently working as an Assistant Professor in the Faculty of Information Technology, Royal University for Women, Bahrain.

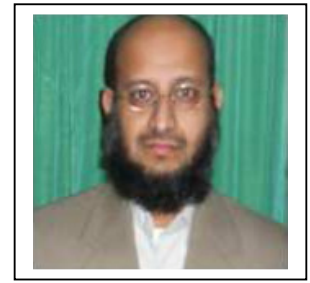